\documentclass[fleqn,useAMS,usenatbib]{mn2e}
\usepackage{graphicx}
\usepackage{amsmath}
\usepackage{amssymb}

\title[Principal Component Analysis of Spectral Line Data]{Principal Component Analysis of Spectral Line Data: Analytic Formulation}
\author[C. M. Brunt]{C. M. Brunt$^{1}$\thanks{E-mail brunt@astro.ex.ac.uk} and M. H. Heyer$^2$ \\
$^{1}$School of Physics, University of Exeter, Stocker Road, Exeter, UK\\
$^{2}$Department of Astronomy, University of Massachusetts, Amherst, MA 01003, USA\\
}
\begin{document}

\date{Accepted ; Received ; in original form }

\pagerange{\pageref{firstpage}--\pageref{lastpage}} \pubyear{2012}

\maketitle

\label{firstpage}

\begin{abstract}
Principal component analysis is a powerful statistical system to investigate the structure and dynamics of the molecular
interstellar medium, with particular emphasis on the study of turbulence, as revealed by spectroscopic imaging of molecular line emission.  
To-date, the method to retrieve the power law index of the velocity structure function or power spectrum has relied on an 
empirical calibration and testing with model turbulent velocity fields, while
lacking a firm theoretical basis. In this paper, we present an analytic formulation
that reveals the detailed mechanics of the method and confirms previous empirical 
calibrations of its recovery of the scale dependence of turbulent velocity 
fluctuations. 
\end{abstract}


\section{Introduction}
Wide field, spectroscopic imaging of molecular line emission provides a vast amount of information of the 
gas dynamics of interstellar clouds.  To exploit this information, Heyer \& Schloerb (1997; HS97) introduced 
the application of Principal component analysis (PCA) to the position-position-velocity data cubes as a tool
to investigate the structure and dynamics of molecular clouds.   
Brunt \& Heyer (2002(a); BH02) more rigorously defined HS97's method for quantifying the
scale-dependence of turbulent velocity fluctuations in molecular clouds, and HS97's
PCA formulation has since undergone a number of extensions and refinements (Brunt 2003(a);
Brunt et al 2003; Heyer et al 2008; Roman-Duval et al 2011). The HS97/BH02 PCA formulation is ideally-suited to
analysis of low signal-to-noise data and for this reason has been most commonly
applied to wide-field survey data (Brunt \& Heyer 2002(b); Heyer \& Brunt 2004; Roman-Duval
et al 2011; Heyer \& Brunt 2012).

A significant limitation of the HS97/BH02 method to derive the power law index of the 
velocity structure function is its reliance on an empirical
calibration that establishes the relationship between the index determined from PCA and the true index of the models 
generated by numerical representations and computational simulations of turbulent
clouds (Brunt \& Heyer 2002a; Brunt et al 2003; Roman-Duval et al 2011).  Therefore, the data analysis has lacked a firm 
theoretical underpinning upon which other statistical methods are based (Scalo 1984; Kleiner \& Dickman 1985; 
Miesch \& Bally 1994; Stutzki et al 1998;  Lazarian \& Pogosyan 2000). 

In this paper, we present an analytic formulation of the PCA method that validates these previous empirical calibrations.
This is a challenging task as it requires analytical representations of a complex physical process
(turbulence) as measured by a complex analysis method (PCA). To simplify the
problem, the formulation relies on a central assumption that the spectral 
line profiles in a spectral line imaging observation of a molecular cloud can be 
represented as an ensemble of Gaussians of fixed dispersion, with turbulent
spatial correlations.  
The formulation predicts covariance matrices, eigenvectors,
eigenvalues, and eigenimage structure and enables insight
into the mechanics of the PCA method that explains several empirically observed
features noted in the literature (Brunt et al 2003; Roman-Duval et al 2011). 

The layout of the paper is as follows. In Section~2 we provide a brief summary of the HS97
formulation. In Section~3 we derive covariance matrices expected from an ensemble of
Gaussian line profiles with variable centroids. Sections~4,~5,~and~6, respectively describe
the derivation of the resultant eigenvectors, eigenvalues, and eigenimages. In Section~7,
we present an analytic derivation of BH02's calibration of the PCA method for the turbulent
velocity fluctuation spectrum. A summary is given in Section~8.

\section{Principal Component Analysis}

In this Section, we review the HS97 formulation of PCA applied to spectral
line imaging observations, and summarise the key empirical findings that an
analytic formulation should aim to explain.

\subsection{The HS97 PCA Formulation}

A spectroscopic imaging observation is comprised of an ensemble of
$n$ spectra each with $p$ spectroscopic channels.
We write the data cube as $T(\mathbfit{r}_{i},v_{j})=T_{ij}$,
where $\mathbfit{r}_{i}$
denotes the spatial coordinate of the $i^{th}$ spectrum.

In the formulation of HS97, the spectrum, or
line profile, at each spatial grid point is
taken to be the raw measurable quantity that
will be subjected to PCA. 
From the ensemble of line profiles, the covariance matrix $S_{jk}$ is calculated as
\begin{equation}
S_{jk} = S(v_{j},v_{k}) = \frac{1}{n} \displaystyle\sum_{i=1}^{n} T_{ij}T_{ik} 
\label{eq:covar}
\end{equation}

A set of
eigenvectors, $u_{mj}$~=~$u_{m}(v_{j})$, and eigenvalues, $\lambda_{m}$, are
determined from the solution of the eigenvalue equation for
the covariance matrix,
\begin{equation}
$$  S_{jk} u_{mj} = {\lambda_{m}} u_{mj} $$
\label{eq:eigvaleq}
\end{equation}
The eigenvalue, $\lambda_{m}$, equals the amount of variance projected
onto its corresponding eigenvector, $u_{mj}$.

The eigenimages, $I_m(\mathbfit{r}_{i})$, are constructed from
the projected values of the
data, $T_{ij}$, onto the eigenvectors, $u_{mj}$,
\begin{equation}
I_{m}(\mathbfit{r}_{i}) = \sum_{j=1}^{p} T_{ij}u_{mj} .
\label{eq:eigimeq}
\end{equation}
We refer to the coupled eigenvector
and eigenimage at order $m$
as the $m^{th}$ principal component (PC). 
In the most basic interpretation, the set of eigenvectors describe 
the velocity magnitude of line profile differences with the ppv volume, as these generate varying levels of variance.   
Such differences arise from gas motions such as infall, outflow, rotation, turbulent velocity fluctuations, and of course, random 
noise of the observation.  The eigenimages show where these profile differences occur within the projected position-position plane.  


\subsection{Empirical Results}

In their foundational work, HS97 suggested that, at each order $m$, the coupled
eigenvector (as a velocity function) and eigenimage (as a spatial function) could
be used to study the scale-dependence of velocity fluctuations in molecular clouds.
Specifically, defining ${\delta}v_{m}$ and ${\delta}l_{m}$ as the characteristic
widths of the eigenvector and eigenimage autocorrelation functions (ACFs) respectively,
HS97 found power-law relations (${\delta}v_{m} \propto {\delta}l_{m}^{\alpha}$)
for a sample of molecular clouds subjected to PCA. 

HS97's proposed method ${\delta}v_{m}$ 
was scrutinised by BH02, who included accounting for noise and finite resolution,
and fixed ${\delta}v_{m}$ and ${\delta}l_{m}$ as the $1/e$ points of the
eigenvector and eigenimage ACFs respectively. BH02 also investigated the method's
ability to recover intrinsic 3-dimensional statistical information about the
velocity field and established the first calibration of the method: $\alpha \approx 0.33\beta$
where $\beta$ is the spectral slope of the angular integral of the velocity power
spectrum in 3D (in this representation, a Kolmogorov spectrum has $\beta = 5/3$ and
a shock-dominated spectrum has $\beta = 2$). Roman-Duval et al (2011) confirmed the
BH02 calibration and examined in detail the sensitivity of the calibration to
density fluctuations, using lognormal density PDFs, concluding that the calibration 
was stable below a critical level of (very high) density 
variability ($\sigma_{\ln{(\rho/\rho_{0})}} > 2$). Brunt et al (2003(a)) and Roman-Duval et al (2011)
showed that the method is sensitive to first-order velocity fluctuations, rather than
root-mean-square velocity fluctuations.

\section{Covariance Matrices}

Our analysis begins with a basic investigation of the covariance matrices that result from an ensemble of Gaussian line profiles of fixed dispersion.
We initially examine the
case of a single component per line of sight, and then consider the more complex case of multiple Gaussians.
This analysis forms the basis of later derivations in the subsequent Sections.

\subsection{Single Gaussian Component Case}

We first consider the covariance matrix that would be derived from an ensemble of Gaussian
line profiles. Let all line profiles have the same dispersion, $\sigma^{2}_{b}$, and
let the distribution of centroid velocities be drawn from a
Gaussian distribution of dispersion $\sigma^{2}_{c}$
around a global mean velocity of zero. The total velocity
dispersion of this ensemble is 
$\sigma^{2}_{tot} = \sigma^{2}_{b} + \sigma^{2}_{c}$.
Note that here, the subscript $b$ refers generically to ``broadening'' of the line profile
due to macroscopic turbulent fluctuations along the line of sight, and not just to the (typically
much narrower) thermal broadening. The use of a single dispersion $\sigma^{2}_{b}$ to represent this
is a simplification, as not all lines of sight will produce exactly the same broadening,
though observationally linewidths do not vary significantly across a cloud.

The terms representing the $i^{th}$ spectrum, $T_{ij}$ and $T_{ik}$, in the covariance matrix 
equation are written:
\begin{equation}
T_{ij} = T_{i}(v_{j}) = T_{0i} \mathrm{exp} \left( - \frac{ (v_{j} - v_{ci})^{2}}{2\sigma^{2}_{b}} \right) ,
\end{equation}
\begin{equation}
T_{ik} = T_{i}(v_{k}) = T_{0i} \mathrm{exp} \left( - \frac{ (v_{k} - v_{ci})^{2}}{2\sigma^{2}_{b}} \right) ,
\end{equation}
where $T_{0i}$ is the peak temperature, $v_{ci}$ is the centroid velocity and $\sigma^{2}_{b}$ is
the velocity dispersion of the $i^{th}$ line profile. 

For the above model, the covariance matrix equation is:
\begin{equation}
S_{jk} = \frac{1}{n} \displaystyle\sum_{i=1}^{n}  T^{2}_{0i} \mathrm{exp} \left( - \frac{ (v_{j} - v_{ci})^{2}}{2\sigma^{2}_{b}} \right) \mathrm{exp} \left( - \frac{ (v_{k} - v_{ci})^{2}}{2\sigma^{2}_{b}} \right) ,
\end{equation}
where the summation is over the total number of line profiles, $n$. 
For large enough $n$ we can convert the
normalised summation over $i$ to integrals over the probability distributions
of peak temperature, $T_{0}$, and centroid velocity, $v_{c}$, to write: 
\begin{eqnarray}
\lefteqn{S_{jk} = \displaystyle\int_{0}^{\infty} dT_{0} \displaystyle\int_{-\infty}^{\infty} dv_{c} P_{T}(T_{0}) P_{v}(v_{c}) T^{2}_{0} } \nonumber \\
\lefteqn{\;\;\;\;\;\;\;\;\;\;\;\;\;\;\;\;\; \times \mathrm{exp} \left( - \frac{ (v_{j} - v_{c})^{2}}{2\sigma^{2}_{b}} \right) \mathrm{exp} \left( - \frac{ (v_{k} - v_{c})^{2}}{2\sigma^{2}_{b}} \right) , }
\label{eq:eq7}
\end{eqnarray}
where we have assumed that $T_{0}$ and $v_{c}$ are uncorrelated, with
independent probability distributions, $P_{T}(T_{0})$ and $P_{v}(v_{c})$ respectively.
Assuming a Gaussian probability distribution for $v_{c}$, with 
dispersion $\sigma^{2}_{c}$, the integrals are easily solved to yield:
\begin{equation}
S_{jk} = S_0 \; \mathrm{exp} \left( - \frac{(v_{j}^{2}+v_{k}^{2})}{2\sigma_{b}^{2}} + \frac{(v_{j}+v_{k})^{2}}{4\sigma_{b}^{2}(1 +  \sigma_{b}^{2}/2\sigma_{c}^{2})} \right) , 
\label{eq:sjkeq}
\end{equation}
where:
\begin{equation}
S_{0} = \frac{\langle T^{2}_{0} \rangle}{\sqrt{1 + 2\sigma^{2}_{c}/\sigma^{2}_{b}}} .
\end{equation}

Equation~(\ref{eq:sjkeq}) is 
valid for ensembles where the peak temperature of the lines can
vary with position, provided the peak temperatures are
uncorrelated with the centroid velocities. Note that the contribution
of a line profile to $S_{jk}$ is proportional to $T_{0}^{2}$. For
consistency, this requires that $\sigma^{2}_{c}$ be defined by:
\begin{equation}
\sigma^{2}_{c} = \frac{ \displaystyle \sum_{i=1}^{n} T^{2}_{0i} v^{2}_{ci}}{\displaystyle \sum_{i=1}^{n} T^{2}_{0i}} = \frac{ \displaystyle \sum_{i=1}^{n} W^{2}_{0i} v^{2}_{ci}}{\displaystyle \sum_{i=1}^{n} W^{2}_{0i}} , 
\end{equation}
where $W_{0i} = \sqrt{2\pi} T_{0i} \sigma_{b}$ is the integrated intensity of the
$i^{th}$ line profile in the above model. Ideally, equation~(\ref{eq:eq7})
would include a probability distribution of $\sigma^{2}_{b}$, but the simplification of a constant 
$\sigma^{2}_{b}$ was necessary to make the integration tractable.

\begin{figure*}
\includegraphics[width=174mm]{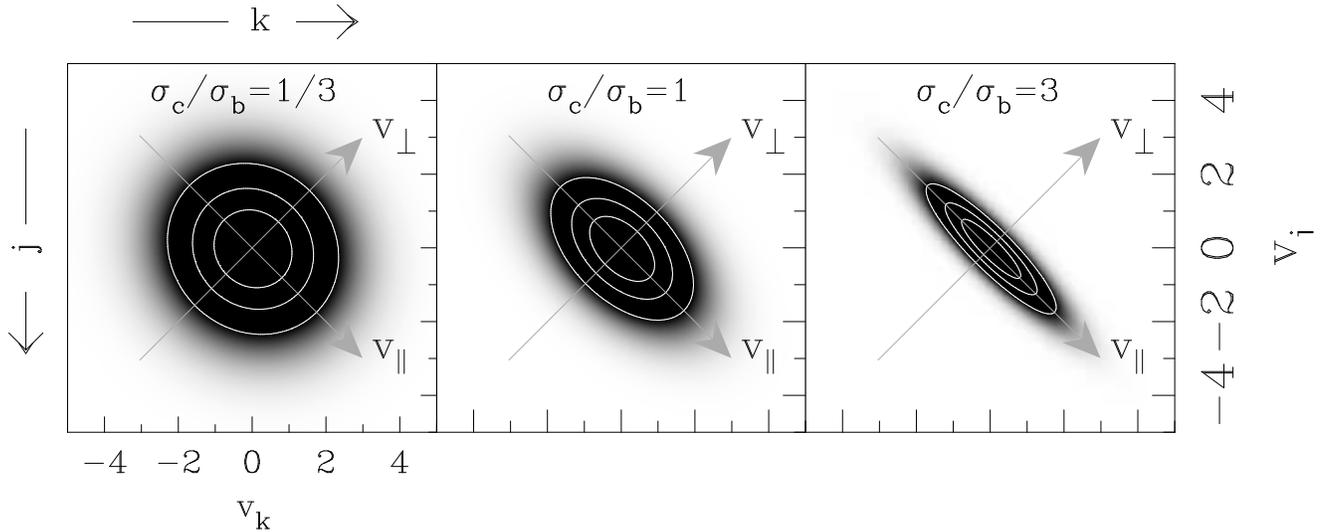}
\caption
{Grayscale representations of the covariance matrix, $S$, obtained with varying $\sigma_{c}/\sigma_{b}$. The variance along the diagonal, $\sigma^{2}_{||} = 2\sigma^{2}_{c} + \sigma^{2}_{b} = 19/9$ is the same for all plots. Contours are shown at 25\%, 50\% and 75\% of the peak of $S_{jk}$.}
\label{fig:sjk}
\end{figure*}

To visualise equation~(\ref{eq:sjkeq}) we constructed covariance matrices according
for varying $\sigma_{b}$ and $\sigma_{c}$. (These matrices agree with numerical
realizations.) Figure~\ref{fig:sjk} shows three
example covariance matrices, represented as
grayscale images. In general, the matrices will vary from
a fully-linearly-dependent case 
($\sigma_{c}$/$\sigma_{b}$~$\longrightarrow$~0)
to a fully-diagonal case 
($\sigma_{b}$/$\sigma_{c}$~$\longrightarrow$~0).
In a fully-diagonal matrix, each row (column) is
linearly independent.

We compute the dispersion of $S$ along
the diagonal, $\sigma^{2}_{||}$, using:
\begin{equation}
S_{||}  =  S_{0}  \mathrm{exp} \left( - \frac{2v_{j}^{2}}{2\sigma_{b}^{2}} + \frac{(2v_{j})^{2}}{4\sigma_{b}^{2}(1 + \sigma_{b}^{2}/2\sigma_{c}^{2})} \right)  = S_{0}  \mathrm{exp} \left( - \frac{2 v_{j}^{2}}{2\sigma_{||}^{2}} \right) , 
\end{equation}
obtained by setting $v_{k} = v_{j}$ in equation~(4), and noting that the distance along the diagonal is $\sqrt{2} v_{j}$, to find:
\begin{equation}
\sigma^{2}_{||} = 2 \sigma_{c}^{2} + \sigma_{b}^{2} .
\end{equation}
Similarly, we compute the dispersion of
$S$ perpendicular to the diagonal, $\sigma^{2}_{\perp}$, using:
\begin{equation}
S_{\perp}  = S_{0}  \mathrm{exp} \left( - \frac{2v_{j}^{2}}{2\sigma_{b}^{2}} \right) = S_{0}  \mathrm{exp} \left( - \frac{2 v_{j}^{2}}{2\sigma_{\perp}^{2}} \right) ,
\end{equation}
obtained by setting $v_{k} = -v_{j}$ in equation~(4), and noting that the distance along the perpendicular is $\sqrt{2} v_{j}$, to find: 
\begin{equation}
\sigma^{2}_{\perp} = \sigma^{2}_{b} .
\end{equation}
More generally, defining:
\begin{equation}
v_{||} = \frac{1}{\sqrt{2}} (v_{k} + v_{j}) ,
\end{equation}
\begin{equation}
v_{\perp} = \frac{1}{\sqrt{2}} (v_{k} - v_{j}) ,
\end{equation}
it is straightforward to show that:
\begin{equation}
S(v_{||},v_{\perp}) = S_{0} \mathrm{exp} \left( - \frac{v^{2}_{\perp}}{2\sigma^{2}_{\perp}} \right) \mathrm{exp} \left( - \frac{v^{2}_{||}}{2\sigma^{2}_{||}} \right) ,
\end{equation}
i.e. that the covariance matrix is an elliptical Gaussian, with dispersions $\sigma^{2}_{||}$ and $\sigma^{2}_{\perp}$ parallel
and perpendicular to the diagonal respectively.

By fitting an elliptical Gaussian to the covariance matrix, $\sigma^{2}_{||}$ and $\sigma^{2}_{\perp}$ can be measured, and we can deduce the line centroid dispersion, $\sigma^{2}_{c}$, and profile dispersion, $\sigma^{2}_{b}$, via:
\begin{equation}
\sigma^{2}_{c} = \frac{1}{2} (\sigma^{2}_{||} - \sigma^{2}_{\perp})
\label{eq:cv1}
\end{equation}
\begin{equation}
\sigma^{2}_{b} = \sigma^{2}_{\perp} .
\label{eq:cv2}
\end{equation} 
It is worth noting also that the total velocity dispersion, $\sigma^{2}_{tot}$, is given by:
\begin{equation}
\sigma^{2}_{tot} = \sigma^{2}_{c} + \sigma^{2}_{b} = \frac{1}{2} (\sigma^{2}_{||} + \sigma^{2}_{\perp}) .
\label{eq:cv3}
\end{equation}

\subsection{Multiple Gaussian Component Case}

We now consider a more elaborate model in which the $i^{th}$ spectrum is represented
by the summation of $n_{t}$ spectral lines, each of dispersion $\sigma^{2}_{t}$, where
we take $n_{t}$ to be moderately large.
Let the centroid velocities of each of these components be drawn from a Gaussian
probability distribution of dispersion $\sigma^{2}_{b} - \sigma^{2}_{t}$ centred
on $v_{ci}$. Here we envision the individual narrow lines to have approximately thermal
linewidths (dispersion $\sigma^{2}_{t}$) that collectively generate a broadened line profile
(with dispersion $\sigma^{2}_{b}$) due to macroscopic velocity differences along the line of sight. 
In the limit of large $n_{t}$, the single component model of the 
preceding section (i.e. a single Gaussian line of dispersion $\sigma^{2}_{b}$
and centroid $v_{ci}$) will be recovered. For moderate $n_{t}$, the line profiles could appear asymmetric 
and/or multiply-peaked, but many profiles averaged together would appear Gaussian.
The contribution of the $i^{th}$ spectrum to the covariance matrix is in this case:
\begin{equation}
\left[\displaystyle \sum_{e=1}^{n_{t}} T_{0ie} \mathrm{exp} \left(- \frac{(v_{j} - v_{cie})^{2}}{2\sigma^{2}_{t}}\right)\right] \times \left[\displaystyle \sum_{f=1}^{n_{t}} T_{0if} \mathrm{exp} \left(- \frac{(v_{k} - v_{cif})^{2}}{2\sigma^{2}_{t}}\right)\right] .
\end{equation}
The contributions for $e = f$:
\begin{equation}
\displaystyle \sum_{e=1}^{n_{t}} T^{2}_{0ie} \mathrm{exp} \left(- \frac{(v_{j} - v_{cie})^{2}}{2\sigma^{2}_{t}}\right) \mathrm{exp} \left(- \frac{(v_{k} - v_{cie})^{2}}{2\sigma^{2}_{t}}\right) 
\end{equation}
averaged over all positions $i$, produce an overall contribution to $S_{jk}$ proportional to:
\begin{equation}
\mathrm{exp} \left( - \frac{(v_{j}^{2}+v_{k}^{2})}{2\sigma_{t}^{2}} + \frac{(v_{j}+v_{k})^{2}}{4\sigma_{t}^{2}(1 +  \sigma_{t}^{2}/2(\sigma_{c}^{2} + \sigma_{b}^{2} - \sigma_{t}^{2}))} \right) ,
\end{equation}
(c.f. Equation~(\ref{eq:sjkeq})). The
contribution of the cross-terms ($e \neq f$) are more difficult to deal with, but
we note that their contribution should recover the form of Equation~(\ref{eq:sjkeq}) in the 
limit of large $n_{t}$. Therefore we write the approximate form of the covariance matrix
in the multiple component case as:
\begin{eqnarray}
S_{jk} \approx \lefteqn{S_{0} \eta \; \mathrm{exp} \left( - \frac{(v_{j}^{2}+v_{k}^{2})}{2\sigma_{t}^{2}} + \frac{(v_{j}+v_{k})^{2}}{4\sigma_{t}^{2}(1 +  \sigma_{t}^{2}/2(\sigma_{c}^{2} + \sigma_{b}^{2} - \sigma_{t}^{2}))} \right)} \nonumber \\
\lefteqn{+ S_{0} (1 - \eta) \; \mathrm{exp} \left( - \frac{(v_{j}^{2}+v_{k}^{2})}{2\sigma_{b}^{2}} + \frac{(v_{j}+v_{k})^{2}}{4\sigma_{b}^{2}(1 +  \sigma_{b}^{2}/2\sigma_{c}^{2})} \right) ,} \nonumber \\
\label{eq:sjkeq2}
\end{eqnarray}
where we expect $\eta \rightarrow 0$ as $n_{t} \rightarrow \infty$. This covariance
matrix form contains an additional (small) contribution from resolvable fine structure
in the line profiles, with dispersion along the diagonal of $2(\sigma^{2}_{c} + \sigma^{2}_{b}) - \sigma^{2}_{t}$
and dispersion perpendicular to the diagonal of $\sigma^{2}_{t}$. Qualitatively, this is a weak,
strongly diagonal feature in the covariance matrix, though this result is obtained only in the large $n_{t}$ limit.

\section{Eigenvectors}

In this Section, we first derive the eigenvectors that result
from a covariance matrix of the form given by Equation~(\ref{eq:sjkeq}). Next,
we derive the autocorrelation functions (ACFs) of the eigenvectors and 
determine the autocorrelation scale, ${\delta}v_{m}$ 
(i.e. the velocity-lag of the $1/e$-point of the normalised ACF) as a 
function of order $m$. This is a key observable in the application of PCA 
to determine the turbulent energy spectrum (HS97; BH02).

\subsection{Eigenvector Structure}

A valid solution of the eigenvalue equation~(\ref{eq:eigvaleq}) requires that:
\begin{equation}
\displaystyle\int_{-\infty}^{+\infty} \; dv_{k} \; S(v_{j},v_{k})  u(v_{k})  =  \lambda  u(v_{j})  ,
\label{eq:eigvalint}
\end{equation}
where $u(v_{k})$ is an eigenvector, $\lambda$ is its eigenvalue, and
we have approximated the finite sums as integrals.
We now search for a valid a solution of equation~(\ref{eq:eigvalint}), using the
form of equation~(\ref{eq:sjkeq}), by setting:
\begin{equation}
u(v_{k})  =  I_{0}  \mathrm{exp} ( - c v_{k}^{2} )  ,
\label{eq:firsteig}
\end{equation}
where $I_{0}$ and $c$ are constants. We use the single component covariance matrix
given by Equation~(\ref{eq:sjkeq}); an analytic solution for the multiple
component case ((Equation~(\ref{eq:sjkeq2})) has not yet been found.

The terms in the exponent of equation~(\ref{eq:sjkeq}) may be written:
\begin{equation}
 - (a  v_{j}^{2}  +  a  v_{k}^{2}  -  2  b  v_{j}  v_{k} )  ,
\label{eq:sub1}
\end{equation} 
where
\begin{equation}
a  = \frac{1}{2\sigma_{b}^{2}} - \frac{1}{4\sigma_{b}^{2}  (1  +  \sigma_{b}^{2}/2\sigma_{c}^{2})}  ,
\end{equation}
and
\begin{equation}
b  = \frac{1}{4\sigma_{b}^{2}  (1 + \sigma_{b}^{2}/2\sigma_{c}^{2})}  .
\end{equation}
The exponent of the integrand in equation~(\ref{eq:eigvalint}) is then:
\begin{equation}
- ( a v_{j}^{2} +  a v_{k}^{2}  -  2 b v_{j} v_{k}  +  c  v_{k}^{2} )  ,
\end{equation} 
which may be regrouped as:
\begin{equation}
- \left[ \left((a + c)^{1/2} v_{k}  -  \frac{b}{(a + c)^{1/2}} v_{j} \right)^{2}  +  \left( a  -  \frac{b^{2}}{(a + c)} \right) v_{j}^{2} \right]  .
\end{equation}
With a change of variable:
\begin{equation}
w  =  (a + c)^{1/2} v_{k}  -  \frac{b}{(a + c)^{1/2}} v_{j} ,
\end{equation}
we find that equation (\ref{eq:eigvalint}): is satisfied if:
\begin{equation}
a - \frac{b^{2}}{(a + c)} = c ,
\end{equation}
or:
\begin{equation}
c^{2}  =  a^{2}  -  b^{2}  = \frac{1}{4} \left( \sigma_{b}^{2}(2\sigma_{c}^{2} + \sigma_{b}^{2}) \right)^{-1/2} .
\label{eq:sub2}
\end{equation}
We identify the solution (equation~(\ref{eq:firsteig})) as the first 
eigenvector ($u_{1j}$~=$~u_{1}(v_{j})$), and  demonstrate 
the validity of this choice below. For simplicity, we write the solution as:
\begin{equation}
u_{1}(v_{j}) =  I_{01}  \mathrm{exp} \left( - \; \frac{v_{j}^{2}}{2\sigma_{1}^{2}} \right)  ,
\end{equation}
where $I_{01}$ is a constant, and: 
\begin{equation}
\sigma_{1}  =  \sqrt{1/2c}  =  (\sigma_{b}^{2}(2\sigma_{c}^{2} + \sigma_{b}^{2}))^{1/4} = \sigma^{1/2}_{||}\sigma^{1/2}_{\perp} .
\label{eq:sig1c}
\end{equation}

\begin{figure*}
\includegraphics[width=84mm]{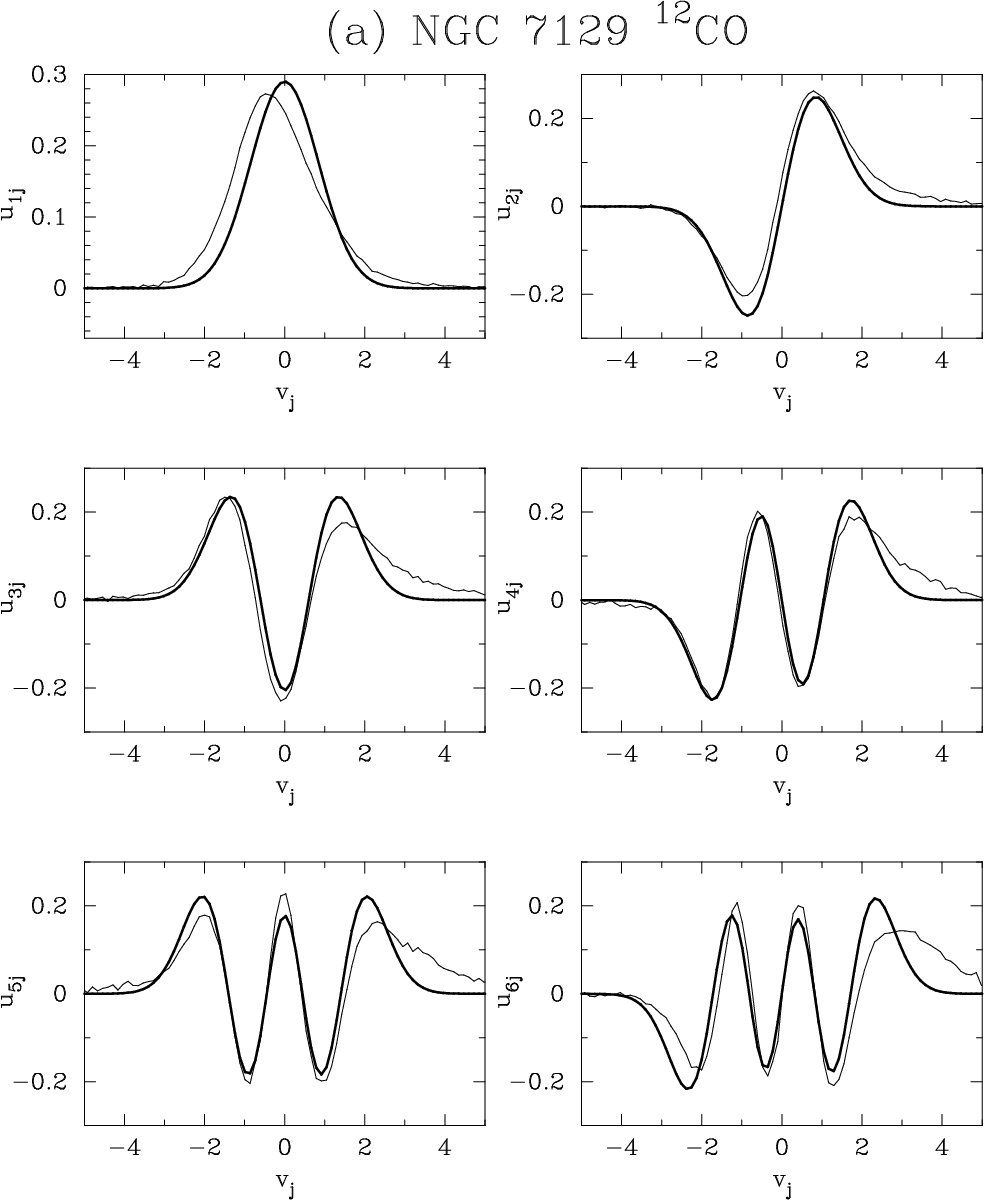}
\includegraphics[width=84mm]{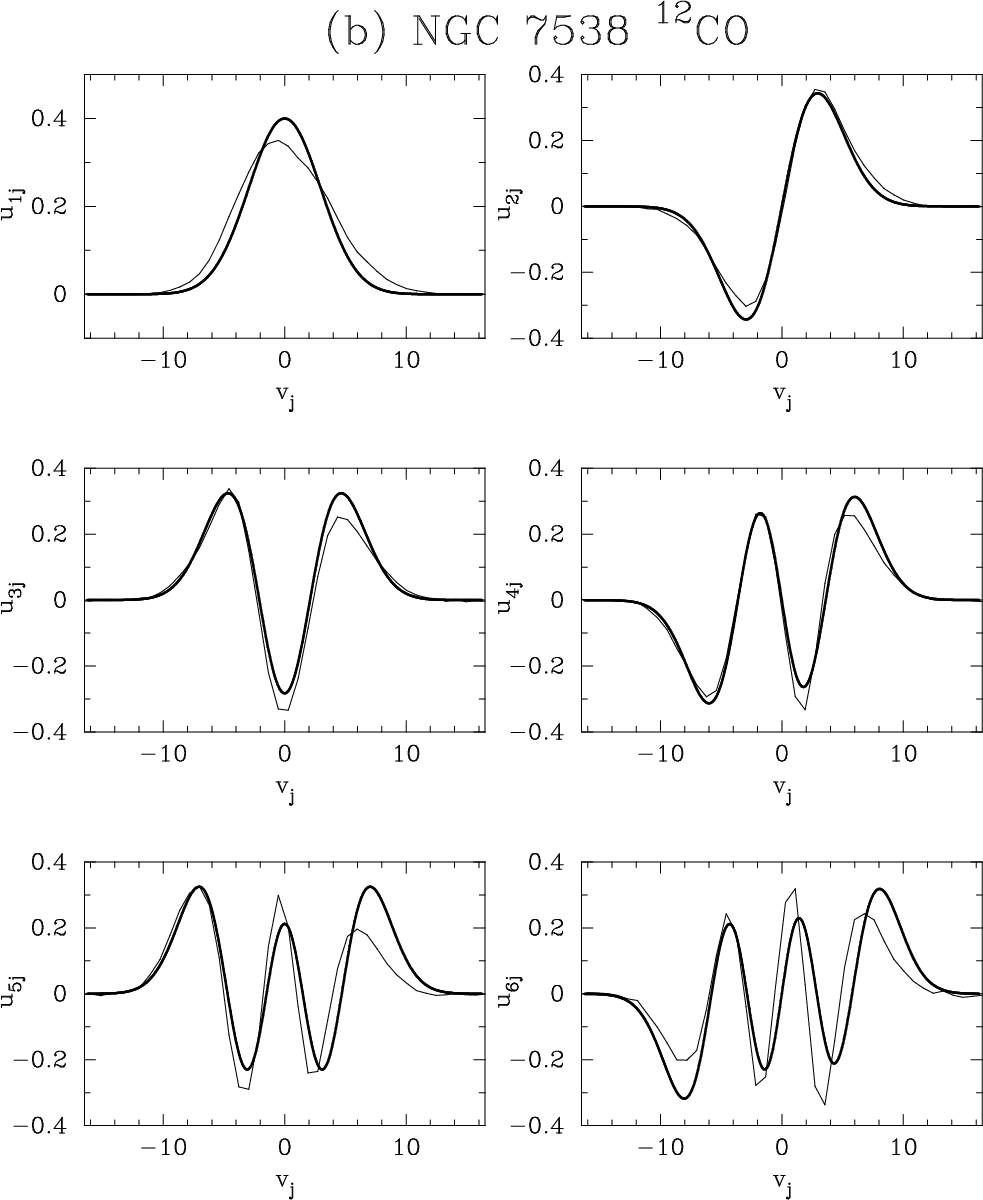}
\caption
{Eigenvectors, $u_{nj} = u_{n}(v_{j})$, obtained from (a)NGC~7129 $^{12}$CO and (b)NGC~7538 $^{12}$CO (lighter lines). Eyeball fits to the fourth eigenvector, $u_{4}$, have been made using the form given in equation~(\ref{eq:analyticeig}). 
The heavy lines are those predicted by equation~(\ref{eq:analyticeig}) with $\sigma_{1}$ and $I_{01}$ specified.}
\label{fig:obseig}
\end{figure*}

To deduce the forms of the higher order eigenvectors, we
make use of the orthogonality condition:
\begin{equation}
\displaystyle\int_{-\infty}^{+\infty} \; dv_{j}  u_{m}(v_{j})  u_{n}(v_{j})  =  I_{0m}  I_{0n}  \delta_{mn} ,
\label{eq:h1}
\end{equation}
where $I_{0m}$ and $I_{0n}$ are constants which depend on the
choice of normalization of the eigenvectors, and $\delta_{mn}$ is the
Kronecker delta ($\delta_{mn} = 1$ if $m = n$, and $\delta_{mn} = 0$ if $m \neq n$). 

The set of functions that are orthogonal with respect to a Gaussian weight are the
Hermite polynomials. The orthogonality condition for Hermite polynomials is:
\begin{equation}
\displaystyle\int_{-\infty}^{\infty} \; dx  H_{n}(x)  H_{m}(x)  \mathrm{exp} (-x^{2})  =  \delta_{mn}2^{n}n!\sqrt{\pi} .
\label{eq:h2}
\end{equation}
Comparing equations (\ref{eq:h1}) and (\ref{eq:h2}), we identify the $m^{th}$ order eigenvector
as the product of the first eigenvector and the $(m-1)^{th}$ order
Hermite polynomial, $H_{m-1}(v_{j}/\sigma_{1})$. 
Thus the $m^{th}$ order eigenvector has the form :
\begin{equation}
u_{mj}  = \frac{I_{01}}{\sqrt{2^{m-1} (m-1)!}}  \mathrm{exp} \left( - \frac{v_{j}^{2}}{2\sigma_{1}^{2}} \right) H_{m-1} \left(\frac{v_{j}}{\sigma_{1}}\right) ,
\label{eq:analyticeig}
\end{equation}
where $I_{01}$ is the peak amplitude of the first eigenvector.

The eigenvectors defined by Equation~(\ref{eq:analyticeig})
provide a reasonably good representation of eigenvectors obtained from
spectral line imaging observations of CO isotopes in molecular clouds. 
Figures~\ref{fig:obseig}(a)(b)
show the first six eigenvectors obtained from
PCA of $^{12}$CO emission in the NGC~7129
molecular cloud (Brunt \& Mac Low 2004)
and the NGC~7538 giant molecular cloud
(Heyer et al. 1998) respectively.
We have fitted (by eye) the fourth eigenvectors with
$u_{4}$ from Equation~(\ref{eq:analyticeig}) and constructed the
other eigenvectors according to $\sigma_{1}$ and
$I_{01}$ obtained from the fit of $u_{4}$. 
The point here is not
to evaluate the detailed applicability of
Equation~(\ref{eq:analyticeig}) to real observations, which
contain more sources of line profile
variance than accounted for by our simple model.
Line profile asymmetries, multiplicities and
other non-Gaussian features will be represented
in the covariance matrix and in turn will
affect the detailed structure of the eigenvectors.
Figure~\ref{fig:obseig} is presented to demonstrate
that observed eigenvectors at order $m$ can be
interpreted as the product of a $\sim$Gaussian
and a polynomial of order $m-1$.

\subsection{Eigenvector Autocorrelation Functions and Characteristic Velocity Scales}{\label{sec:eigenvectorscales}}

\begin{figure*}
\includegraphics[width=84mm,angle=-90]{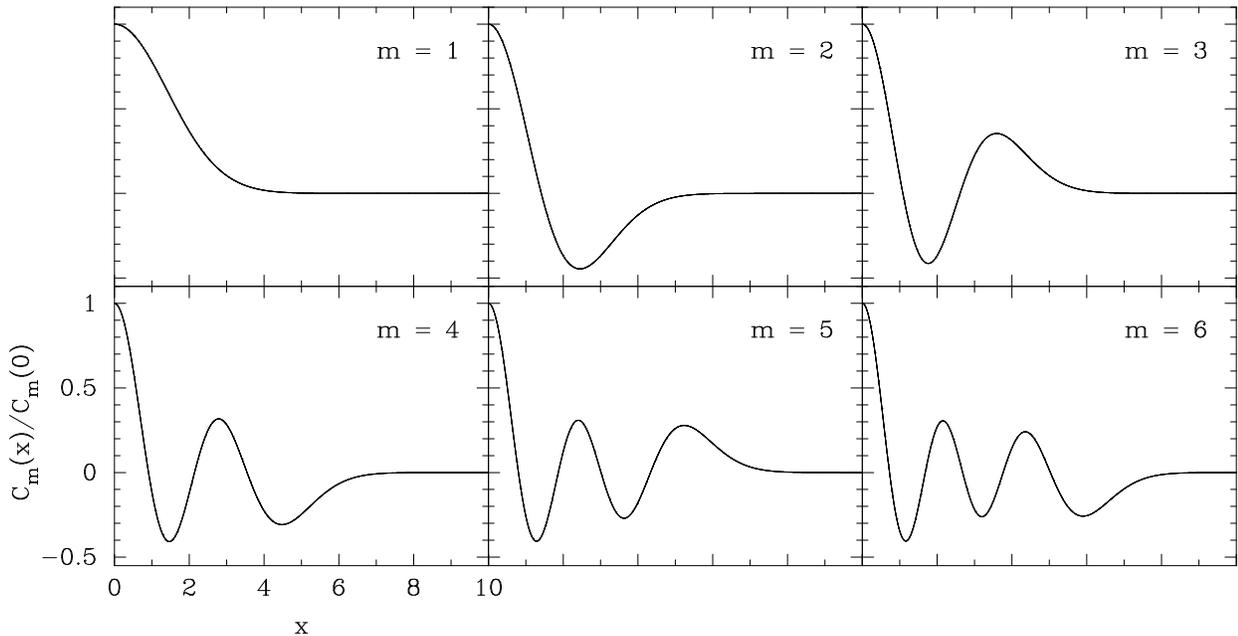}
\caption{The first six eigenvector ACFs, given by equation~(\ref{eq:eigacfeq}).}
\label{fig:eigacfplot}
\end{figure*}

The unnormalised autocorrelation function (ACF), $C_{m}(v)$ of the $m^{th}$ eigenvector is:
\begin{equation}
C_{m}(v) = \displaystyle \int_{-\infty}^{\infty} dv' u_{m}(v') u_{m}(v' - v) ,
\end{equation}
where $u_{m}(v)$ at order $m$ is given by Equation~(\ref{eq:analyticeig}).
Writing $x = v/\sigma_{1}$ and $y = v'/\sigma_{1}$, this is then:
\begin{eqnarray}
\lefteqn{C_{m}(x) = C_{m}(v/\sigma_{1})} \nonumber \\
 \lefteqn{= \frac{I_{01}^{2}}{2^{m-1} (m-1)!} \displaystyle \int_{-\infty}^{\infty} dy \; \mathrm{exp} \left( -y^{2}/2 \right)} \nonumber \\
\lefteqn{ \;\;\;\;\;\;\;\;\;\;\;\;\; \times  \mathrm{exp} \left( -(x - y)^{2}/2 \right) H_{m-1}(y) H_{m-1}(y-x) .}  \nonumber \\
\end{eqnarray}
We make the substitution $w = y - x/2$ to find:
\begin{eqnarray}
\lefteqn{C_{m}(x) = C_{m}(v/\sigma_{1})} \nonumber \\
 \lefteqn{= \frac{I_{01}^{2}}{2^{m-1} (m-1)!} \displaystyle \int_{-\infty}^{\infty} dw \; \mathrm{exp} \left( -w^{2}
 \right)} \nonumber \\
\lefteqn{ \;\;\;\;\;\;\;\;\;\;\;\;\; \times  \mathrm{exp} \left( -(x/2)^{2} \right) H_{m-1}(w+x/2) H_{m-1}(w-x/2) .}  \nonumber \\
\end{eqnarray}
The Hermite polynomial terms may be expanded as:
\begin{eqnarray}
\lefteqn{H_{m-1}(w+x/2) = \displaystyle \sum_{k=0}^{m-1} \frac{(m-1)!}{k!(m-1-k)!} H_{k}(w) x^{m-1-k} ,} \nonumber \\
\lefteqn{H_{m-1}(w-x/2) = \displaystyle \sum_{k=0}^{m-1} \frac{(m-1)!}{k!(m-1-k)!} H_{k}(w) (-x)^{m-1-k} .} \nonumber \\
\end{eqnarray}
Using the orthogonality of Hermite polynomials (Equation~(\ref{eq:h2})), this then gives:
\begin{equation}
\frac{C_{m}(x)}{C_{m}(0)} = \frac{C_{m}(v/\sigma_{1})}{C_{m}(0)} = \mathrm{exp} \left( -(x/2)^{2}\right) B_{m-1}(x),
\label{eq:eigacfeq}
\end{equation}
where:
\begin{eqnarray}
\lefteqn{B_{m-1}(x) = } \nonumber \\
\lefteqn{\displaystyle \sum_{k=0}^{m-1} \frac{2^{-(m-1-k)}}{(m-1-k)!} \frac{(m-1)!}{k!(m-1-k)!} (-1)^{m-1-k} x^{2(m-1-k)} . } \nonumber \\
\end{eqnarray}
Note that we have also written these in normalised form.

The first five normalised ACFs are:
\begin{eqnarray}
\lefteqn{\frac{C_{1}(x)}{C_{1}(0)} = \mathrm{exp} \left( -(x/2)^{2}\right)} \nonumber \\
\lefteqn{\frac{C_{2}(x)}{C_{2}(0)} = \mathrm{exp} \left( -(x/2)^{2}\right) (1 - x^{2}/2)} \nonumber \\
\lefteqn{\frac{C_{3}(x)}{C_{3}(0)} = \mathrm{exp} \left( -(x/2)^{2}\right) (1 - x^{2} + x^{4}/8)} \nonumber \\
\lefteqn{\frac{C_{4}(x)}{C_{4}(0)} = \mathrm{exp} \left( -(x/2)^{2}\right) (1 - 3x^{2}/2 + 3x^{4}/8 - x^{6}/48)} \nonumber \\
\lefteqn{\frac{C_{5}(x)}{C_{5}(0)} = \mathrm{exp} \left( -(x/2)^{2}\right) (1 - 2x^{2} + 3x^{4}/4 - x^{6}/12 + x^{8}/384) .} \nonumber \\
\end{eqnarray}
The first six ACFs are shown in Figure~\ref{fig:eigacfplot} -- c.f. Figure~9 of HS97.

\begin{figure}
\includegraphics[width=84mm]{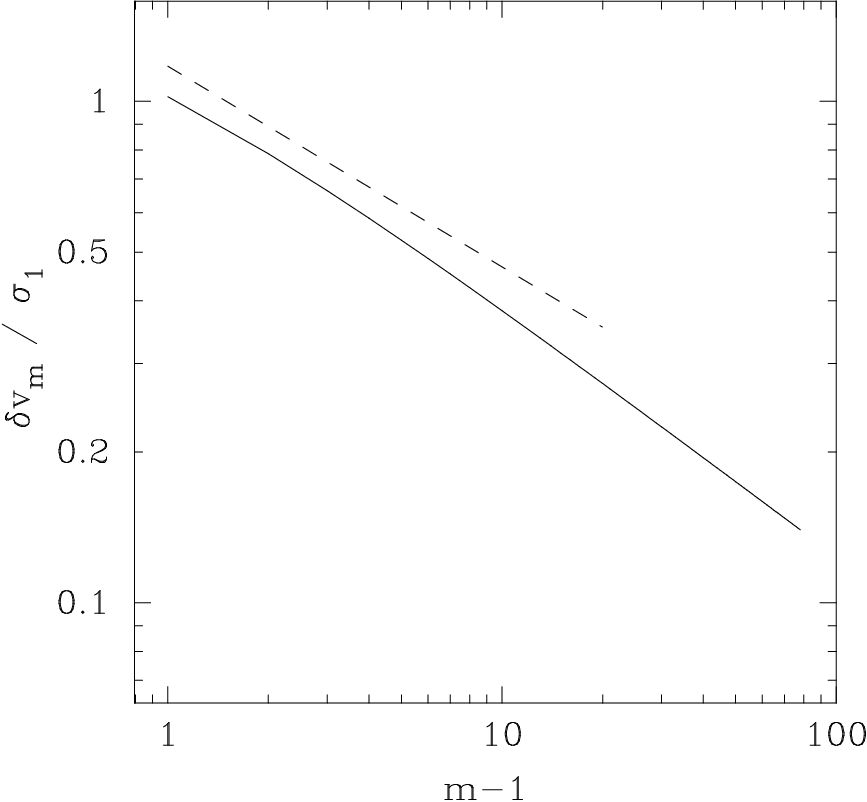}
\caption{Log-log plot of the measured values of ${\delta}v_{m}/\sigma_{1}$ determined at the $1/e$ points of the eigenvector ACFs versus $m-1$. For
reference, the dashed line (offset) has a slope of $-\xi = -0.4$, appropriate for low orders $m$.}
\label{fig:dvm}
\end{figure}

\begin{figure}
\includegraphics[width=84mm]{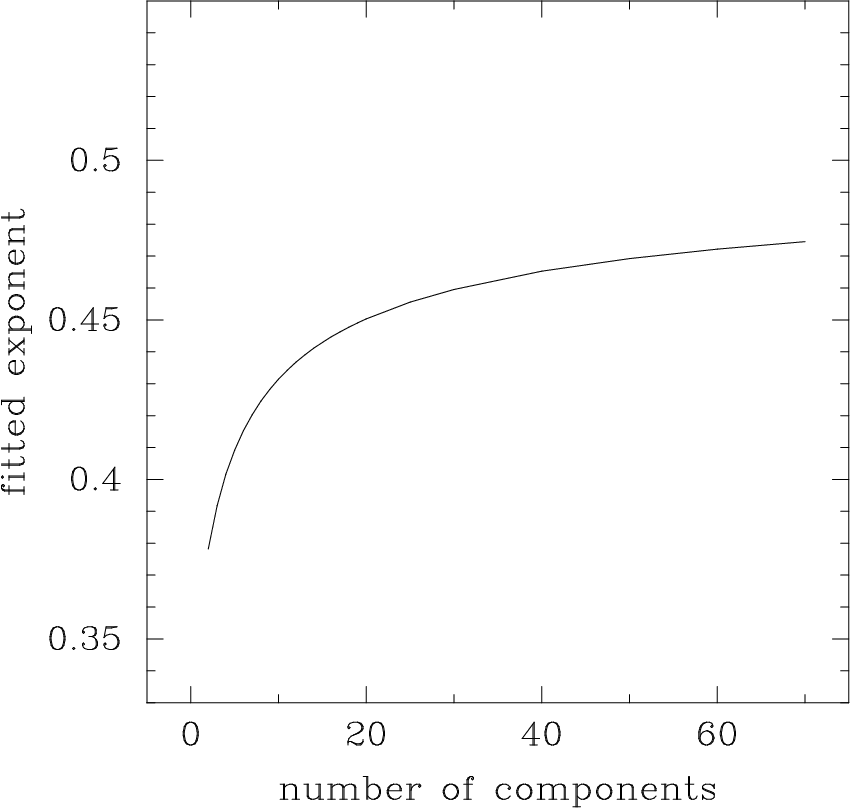}
\caption{The fitted exponent, $\xi$, from equation~(\ref{eq:dvscales}) as a function of the number of recovered components from which the fit is made.}
\label{fig:bmxslope}
\end{figure}

The velocity scale, ${\delta}v_{m}$, at order $m$ is given by the $1/e$-point of the normalised
ACF, i.e. $C_{m}({\delta}v_{m}/\sigma_{1})/C_{m}(0) = 1/e$. 
While it is difficult to determine the $1/e$ points analytically, they may
be determined numerically. Figure~\ref{fig:dvm} shows the measured ${\delta}v_{m}/\sigma_{1}$ values
versus $m-1$, which approximately obey a power law relation:
\begin{equation}
{\delta}v_{m}/\sigma_{1} \propto (m-1)^{-\xi} .
\label{eq:dvscales}
\end{equation}
However, closer inspection reveals that in practice the exponent $\xi$ is dependent on the maximum
number of recovered components. In Figure~\ref{fig:bmxslope} we plot the fitted exponent, $\xi$, as a function
of the number of recovered components. For only two recovered components, $\xi \approx 0.38$, while in the
(practically unachievable) limit of a very large number of recovered components, $\xi$ asymptotically approaches 0.5. 
For a representative number of recovered components (between 3 and 20) in the calibration data of BH02,
we adopt a working value of $\xi = 0.4\pm0.02$.

\section{Eigenvalues}

\begin{figure}
\includegraphics[width=84mm]{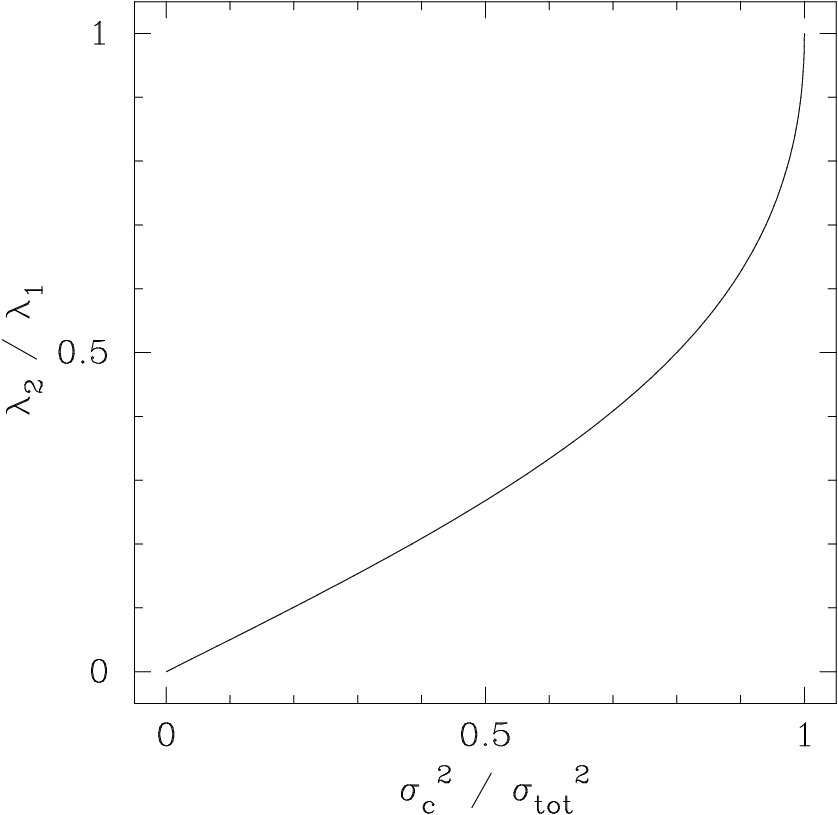}
\caption{Relation between $\sigma^{2}_{c}/\sigma^{2}_{tot}$ and the ratio of the first two eigenvalues $\lambda_{2} / \lambda_{1}$.}
\label{fig:eigenvalue}
\end{figure}

For eigenvectors given by equation~\ref{eq:analyticeig}, it is possible to deduce
the corresponding eigenvalues using equation~\ref{eq:eigvalint}. 
For the first two eigenvectors, equation~\ref{eq:eigvalint} reads:
\begin{equation}
\displaystyle\int_{-\infty}^{+\infty} \; dv_{k} \; S(v_{j},v_{k}) I_{01} \mathrm{exp} \left( - \frac{v_{k}^{2}}{2\sigma_{1}^{2}} \right)  =  \lambda_{1} I_{01} \mathrm{exp} \left( - \frac{v_{j}^{2}}{2\sigma_{1}^{2}} \right) ,
\label{eq:sus1}
\end{equation}
\begin{eqnarray}
\lefteqn{\displaystyle\int_{-\infty}^{+\infty} \; dv_{k} \; S(v_{j},v_{k}) I_{01} \sqrt{2} \frac{v_{k}}{\sigma_{1}} \mathrm{exp} \left( - \frac{v_{k}^{2}}{2\sigma_{1}^{2}} \right)} \nonumber \\
\lefteqn{  =  \lambda_{2} I_{01} \sqrt{2} \frac{v_{j}}{\sigma_{1}} \mathrm{exp} \left( - \frac{v_{j}^{2}}{2\sigma_{1}^{2}} \right) ,}
\label{eq:sus2}
\end{eqnarray}
where $S(v_{j},v_{k})$ is given by equation~\ref{eq:sjkeq}. Making use of 
equations~(\ref{eq:sub1}--\ref{eq:sub2}), these can be solved to find:
\begin{equation}
\lambda_{1} = \sqrt{\frac{\pi}{a+c}} S_{0},
\end{equation}
\begin{equation}
\lambda_{2} = \sqrt{\frac{\pi}{a+c}} \left( \frac{b}{a+c} \right) S_{0} ,
\end{equation}
which leads to:
\begin{equation}
\frac{\lambda_{2}}{\lambda_{1}} = \frac{\sigma^{2}_{tot}}{\sigma^{2}_{c}} - \sqrt{\left(\frac{\sigma^{2}_{tot}}{\sigma^{2}_{c}}\right)^{2} - 1} ,
\label{eq:eigv1}
\end{equation}
or:
\begin{equation}
\frac{\sigma^{2}_{c}}{\sigma^{2}_{tot}} = \frac{2(\lambda_{2}/\lambda_{1})}{1 + (\lambda_{2}/\lambda_{1})^{2}} .
\label{eq:eigv2}
\end{equation}
Equations~(\ref{eq:eigv1})~and~(\ref{eq:eigv2}), graphically represented in Figure~\ref{fig:eigenvalue}, show that, in the
case of no centroid variation, all the variance of the data is contained in the first (and only) principal component. The maximum
value of $\lambda_{2}/\lambda_{1} = 1$ is found in the limit where all variance in the data is caused by centroid variations. In general,
the ratio $\lambda_{2}/\lambda_{1}$ can be used to provide a straightforward measurement of the ratio $\sigma^{2}_{c}/\sigma^{2}_{tot}$.

\section{Eigenimages}

The covariance matrix and eigenvectors are independent of the spatial structure
of the spectral line data. However, each eigenvector has an associated spatial
map, the ``eigenimage'', formed by projection of the data onto the eigenvector
via Equation~(\ref{eq:eigimeq}). This can be alternatively viewed as the
integration of the data over the velocity axis with the eigenvector
acting as a weighting or windowing function. For low order eigenvectors, there
is a straightfoward interpretation of this procedure and it is possible to
relate the resulting eigenimages to physical measures of the medium from which
the line profiles originate. Below, we first derive the eigenimage structures for the
two lowest order eigenvectors. Following this, we derive the asymptotic form
of the eigenimages at high order.

\subsection{Eigenimage Structure}

The form of the eigenimages, given by Equation~(\ref{eq:eigimeq}), is:
\begin{eqnarray}
\lefteqn{I_{m}({\mathbfit{r}}) = \frac{I_{01}}{\sqrt{2^{m-1} (m-1)!}} \displaystyle \int_{-\infty}^{\infty} dv \; T({\mathbfit{r}}, v) } \nonumber \\
\lefteqn{\;\;\;\;\;\;\;\;\;\;\;\;\;\;\;\;\; \times \mathrm{exp}  \left( - \frac{v^{2}}{2\sigma^{2}_{1}} \right) H_{m-1}(v/\sigma_{1}) , } \nonumber \\
\end{eqnarray}
which can be interpreted as a generator of moments over the brightness temperature,
subject to an overall windowing function $I_{01} \mathrm{exp} (- v^{2}/2\sigma^{2}_{1}) = u_{1}(v)$.

The first few Hermite polynomials are:
\begin{eqnarray}
\lefteqn{H_{0}(x) = 1} \nonumber \\
\lefteqn{H_{1}(x) =  2x} \nonumber \\
\lefteqn{H_{2}(x) =  4x^{2} - 2} \nonumber \\
\lefteqn{H_{3}(x) =  8x^{3} - 12x} \nonumber \\
\lefteqn{H_{4}(x) =  16x^{4} - 48x^{2} + 12 ,} \nonumber \\
\end{eqnarray}
so that the first two eigenimages are closely related to the
$0^{th}$ and $1^{st}$ moments -- i.e. the integrals of $T(v)$
and $T(v)v$ respectively -- but with the additional velocity-windowing
provided by $u_{1}(v)$. 
Brunt (2003(a)) and Brunt, Heyer and Mac Low (2009)
have made use of this to probe the outer scale of turbulence in molecular
clouds, since the $0^{th}$ moment and $1^{st}$ moment are proportional
to the column density and the projected momentum respectively 
(see e.g. Brunt \& Federrath 2013, submitted), with both subject to the same windowing function.

Writing $T({\mathbfit{r}}, v) = T_{0}({\mathbfit{r}}) \mathrm{exp} (-(v - v_{c}({\mathbfit{r}}))^{2}/2\sigma^{2}_{b})$
and choosing the convenient normalisation $I_{01} = 1$, the first
eigenimage, $I_{1}({\mathbfit{r}})$ is:
\begin{equation}
I_{1}({\mathbfit{r}}) = \sqrt{2\pi}\sigma_{b}T_{0}({\mathbfit{r}}) F(v_{c}) ,
\end{equation}
where:
\begin{equation}
F(v_{c}) = F^{1/2}_{0} \mathrm{exp} \left(- \frac{v^{2}_{c}}{2(\sigma^{2}_{1}+\sigma^{2}_{b})} \right) 
\end{equation}
is the integrated effect of the windowing function (with
$F_{0} = \sigma^{2}_{1}/(\sigma^{2}_{1}+\sigma^{2}_{b}))$. Note that
$\sqrt{2\pi}\sigma_{b}T_{0}({\mathbfit{r}}) = W_{0}({\mathbfit{r}})$ is the integrated intensity
($0^{th}$ moment) of the emission. The effect of $F(v_{c})$ is
to attenuate the eigenimage intensity for line profiles 
with high $|v_{c}|$.

The second eigenimage, $I_{2}({\mathbfit{r}})$, is:
\begin{eqnarray}
\lefteqn{I_{2}({\mathbfit{r}}) = \sqrt{2\pi}\sigma_{b}T_{0}({\mathbfit{r}}) v_{c}({\mathbfit{r}}) \frac{F^{3/2}_{0}}{\sqrt{2}\sigma_{1}} F(v_{c})} \nonumber \\
\lefteqn{\;\;\;\;\;\;\;\;\;\;\;\; = \frac{F_{0}}{\sqrt{2}\sigma_{1}} I_{1}({\mathbfit{r}}) v_{c}({\mathbfit{r}}),} \nonumber \\
\label{i2full}
\end{eqnarray}
which is seen to be the $1^{st}$ moment of the intensity again subject
to the integrated effect of the windowing function.

Higher order eigenimages combine
higher order moments, again with windowing by $u_{1}(v)$, but become
increasingly difficult to interpret except in a statistical way.
An approximate form for higher order eigenimages may be arrived
at by making use of the following expansion at high $n$:
\begin{equation}
\mathrm{exp} \left( - \frac{x^{2}}{2}\right) H_{n}(x) \approx \frac{2^{n}}{\sqrt{\pi}} \Gamma\left(\frac{n+1}{2}\right) \mathrm{cos}\left(x\sqrt{2n} - n\frac{\pi}{2}\right) .
\end{equation}
Inserting this expression into Equation~(\ref{eq:eigimeq}), yields,
after some manipulation:
\begin{eqnarray}
\lefteqn{I_{m}({\mathbfit{r}}) \approx G(m) \sqrt{2\pi} \sigma_{b} T_{0}({\mathbfit{r}}) \mathrm{cos}\left(\frac{v_{c}({\mathbfit{r}})}{\sigma_{1}} \sqrt{2(m-1)}\right)  {\mathrm{for~odd~m}} ,} \nonumber \\
\lefteqn{I_{m}({\mathbfit{r}}) \approx G(m) \sqrt{2\pi} \sigma_{b} T_{0}({\mathbfit{r}}) \mathrm{sin}\left(\frac{v_{c}({\mathbfit{r}})}{\sigma_{1}} \sqrt{2(m-1)}\right)  {\mathrm{for~even~m}} ,} \nonumber \\
\label{i2approx}
\end{eqnarray}
where $G(m)$ is an unimportant (constant) $m$-dependent multiplicative factor.
While strictly only accurate at high $m$, these expressions provide a 
reasonably good representation of the eigenimage
structure even at the lowest $m$-values (though quantitaively, the differences are important as
we discuss in the next Section).
Note that for small $v_{c}/\sigma_{1}$, both Equation~(\ref{i2full}) and 
Equation~(\ref{i2approx}) give $I_{2} \propto T_{0}v_{c}$. In addition, the windowing
term, $F_{v_{c}}$, in Equation~(\ref{i2full}) crudely approximates the roll-off in $I_{2}$ caused
by the sinusoidal behaviour in Equation~(\ref{i2approx}). 

The structure of the eigenimages predicted by Equation~(\ref{i2approx}) is
as follows. The overall amplitude (at any order $m$) is controlled by the
column density ($\sqrt{2\pi} \sigma_{b} T_{0}({\mathbfit{r}})$), and this is modulated by
a common multiplicative factor (dependent on $m$) and, more importantly,
a sine or cosine factor, dependent on the centroid velocity, $v_{c}({\mathbfit{r}})$. Therefore,
as the order $m$ increases, the eigenimage values cycle through a sine 
or cosine variation. This provides the key to understanding their characteristic
spatial scale lengths needed for the measurement of the turbulent
velocity spectrum, as described in the next Section.
 
\section{Analytic Calibration of the PCA Method for the Turbulent Velocity Spectrum}

Our procedure here is to generate a coupled sequence of characteristic
spatial and velocity scales (${\delta}l_{m}, {\delta}v_{m}$) at order $m$, for
a specified spectral index $\beta$ of the 3D velocity field. The
dependence of the predicted exponent $\alpha$ (where ${\delta}v_{m} \propto {\delta}l_{m}^{\alpha}$)
on the intrinsic $\beta$ will then establish the calibration (see Section 2.2).

We have already established the $m$-dependence of ${\delta}v_{m}$ in 
Section~\ref{sec:eigenvectorscales}, where it was found that
${\delta}v_{m}$~$\propto$~$(m-1)^{-\xi}$ with $\xi \approx 0.4$. It still
remains to determine the corresponding sequence ${\delta}l_{m}$. Here,
however, while we have a functional form for the asymptotic
eigenimage structure (Equation~(\ref{i2approx})) we do not
have a definite expression for the field $v_{c}({\mathbfit{r}})$, but instead
only have a statistical knowledge of its properties, which may
be quantified via structure functions.

The $p^{th}$-order structure function of a velocity field is
written:
\begin{equation}
S_{p}(l) = \langle |{\Delta}v(l)|^{p} \rangle \propto l^{\zeta_{p}} ,
\end{equation}
where ${\Delta}v(l)$ represents the ensemble of velocity fluctuations 
measured on spatial scale $l$ in the field, and angle brackets denote 
spatial averaging. The function $\zeta_{p}$ describes the dependence 
of the scaling exponent on the order $p$. Alternatively, one may write:
\begin{equation}
(S_{p}(l))^{1/p} = \langle |{\Delta}v(l)|^{p} \rangle^{1/p} \propto l^{\gamma_{p}} ,
\end{equation}
where $\gamma_{p} = \zeta_{p}/p$. For velocity fields produced by
fBm, $\gamma_{p}$ is independent of $p$ (e.g. Brunt et al 2003).
For now, we will assume that the centroid velocity field, $v_{c}(x, y)$
can be described by a scaling exponent $\gamma_{c}$ (valid at all $p$),
allowing us to write:
\begin{equation}
\langle |{\Delta}v_{c}(l)/\sigma_{1}|^{p} \rangle^{1/p} = (l/l_{1})^{\gamma_{c}} ,
\label{eigscaling}
\end{equation}
where $l$ is the 2D spatial scale and and $l_{1}$ is the spatial scale
corresponding to a mean velocity fluctuation of $\sigma_{1}$.

The original calibration established by BH02 used uniform density
fields (and therefore uniform column density fields) so that only the effect of the
(co)sine term in Equation~(\ref{i2approx}) need be inspected. 
The (co)sine term leads to an oscillatory eigenimage structure with a characteristic
spatial wavelength $L_{m}$ set by the condition that {\it the typical velocity fluctuation
between points separated by a distance $L_{m}$ generates a phase difference of
$2\pi$ in the argument of the (co)sine term}. That is:
\begin{equation}
\sqrt{2(m-1)}\langle{\Delta}v_{c}(L_{m})\rangle/\sigma_{1} \approx 2\pi .
\end{equation}
Referring to equation~(\ref{i2approx}), note that
because the $\sqrt{2(m-1)}$ factor effectively {\it amplifies} 
the $v_{c}$ field, progressively smaller velocity fluctuations are
capable of inducing a $2\pi$ phase difference as the order $m$ increases
(i.e. the typical $v_{c}$ fluctuation required falls proportionally
to $(m-1)^{-1/2}$). Consequently, there is a corresponding reduction
in the characteristic spatial wavelength, governed by Equation~(\ref{eigscaling}),
such that:
\begin{equation}
L_{m}/l_{1} \approx \langle |{\Delta}v_{c}(L_{m})/\sigma_{1}| \rangle^{1/\gamma_{c}} \approx \left(\frac{\sqrt{2}\pi}{(m-1)^{1/2}}\right)^{1/\gamma_{c}} ,
\end{equation}
meaning that the characteristic wavelength of eigenimage structure
decreases with order $m$ as $L_{m} \propto (m-1)^{-1/2\gamma_{c}}$.

The characteristic spatial scale, ${\delta}l_{m}$, of the $m^{th}$-order
eigenimage is determined by the $1/e$ point of the eigenimage autocorrelation
function, and it is straightforward to show that for a (co)sinusoid:
\begin{equation}
{\delta}l_{m} = \left(\frac{{\mathrm{acos}}(1/e)}{2\pi}\right) L_{m} \approx 0.19 L_{m} .
\end{equation}
Therefore, the $m$-dependence of characteristic eigenimage scales, in the asymptotic approximation, is:
\begin{equation}
{\delta}l_{m} \propto (m-1)^{-1/2\gamma_{c}} ,
\label{eq:dlscales}
\end{equation}
where $\gamma_{c}$ is the scaling exponent of the centroid velocity field.
However, this is slightly inaccurate as the asymptotic expansions are not
strictly applicable at low order $m$. We note first that, crudely approximating the
$v_{c}$ field as a $\sim$~linear gradient, the exact equation~(\ref{i2full}) predicts
a scale ${\delta}l_{2}$ that is 20\% larger than that predicted by equation~(\ref{i2approx}).
Since as the order $m$ increases, the asymptotic formula becomes increasingly
more accurate, this in effect means that ${\delta}l_{m}$ falls faster with $m-1$
than equation~(\ref{eq:dlscales}) predicts. Assuming a smooth transition between
a $\sim$~20\% overestimation at low $m$ to accurate representation at, say, $m \gtrsim 10$,
we estimate that the effective $m$-dependence of ${\delta}l_{m}$ is better represented by:
\begin{equation}
{\delta}l_{m} \propto (m-1)^{-1.1/2\gamma_{c}} ,
\label{eq:dlscalesfixed}
\end{equation}
i.e. an increase of the exponent, by a factor of 1.1 ($\pm$0.03),
describing the reduction of characteristic spatial scale as the order increases. 

Combining equation~(\ref{eq:dlscalesfixed}) with the $m$-dependence of the characteristic velocity
scales (Equation~(\ref{eq:dvscales})), we arrive at a calibration of the 
PCA $\alpha$ exponent to the centroid velocity scaling exponent, $\gamma_{c}$, 
via:
\begin{equation}
{\delta}v_{m} \propto {\delta}l_{m}^{\alpha} \propto {\delta}l_{m}^{2 \xi \gamma_{c}/1.1} ,
\end{equation}
so that:
\begin{equation}
\alpha \approx 2 \xi \gamma_{c} / 1.1,
\label{eq:dvlcal1}
\end{equation}
and taking the representative value $\xi = 0.4\pm0.02$, as discussed in Section~4.2, this
leads to:
\begin{equation}
\alpha \approx 0.72 \gamma_{c} .
\label{eq:dvlcal2}
\end{equation}

It remains to relate $\gamma_{c}$ to the spectral index, $\beta$, of the 3D
velocity field. This is a general question (not restricted to the PCA method)
but one that has a simple answer in the uniform density conditions assumed
by BH02 in the original calibration. As explained in Brunt \& Mac Low (2004; and
references therein), the following relation holds for uniform density and optically-thin conditions:
\begin{equation}
\gamma_{c} = \frac{\beta}{2} .
\label{gammacbeta}
\end{equation}
Some discussion of this equation is warranted, as the scaling exponent of
the velocity field in 3D ($\gamma_{3D}$, here assumed independent of $p$, appropriate
for the non-intermittent fBm fields used by BH02) is given by:
\begin{equation}
\gamma_{3D} = \frac{\beta - 1}{2} ,
\end{equation}
and therefore:
\begin{equation}
\gamma_{c} = \gamma_{3D} + \frac{1}{2} = \frac{\beta}{2}.
\end{equation}
The increase in the exponent upon projection (by $1/2$) is known as ``projection
smoothing'', and can be qualitatively understood by considering that large-scale
velocity fluctuations suffer proportionally less line-of-sight averaging
than small-scale fluctuations.

Using equations~(\ref{gammacbeta})~and~(\ref{eq:dvlcal2}) we arrive at the analytic calibration of the
PCA $\alpha$ exponent:
\begin{equation}
\alpha \approx 0.36 \beta .
\label{analyticcal}
\end{equation}
This relation is close to, though slightly steeper than, the empirically-determined
$\alpha \approx( 0.33{\pm}0.04)\beta$ (BH02; Roman-Duval et al 2011). This is encouraging
analytic support for the empirical calibration, and the small difference in exponent
(0.36$\pm$0.04 versus 0.33$\pm$0.04) is not too concerning, given the approximations used in the derivations
above. 

In the above, we have not explicitly included the effects of opacity, and it is worth considering how
this may affect the result. Previously, it has been found empirically that opacity/saturation does not
have a drastic affect on $\alpha$ (Brunt et al 2003; Roman-Duval et al 2011). It is also observed that
application of the method to $^{12}$CO and $^{13}$CO data on the same cloud yields very similar 
${\delta}v(\ell)$ spectra and similar values of $\alpha$
(e.g. Brunt 2003(b); Brunt \& Mac Low 2004; Brunt et al 2009). A likely reason for this insensitivity is that
the centroid velocity field is not strongly affected by saturation {\it if} the saturation is symmetric about line
centre. Brunt \& Mac Low (2004) demonstrate directly that the centroid fields derived in their observations
from $^{12}$CO and $^{13}$CO are almost indistinguishable statistically. A secondary effect of saturation
may be to move the line profiles to a flat-topped appearance, invalidating the gaussian form assumed above. However,
the requirement of orthogonality in the eigenvectors essentially ensures a polynomial sequence similar to the derived
Hermite polynomials, so any deviations from our scaling result will likely be small. However, we cannot analytically
assess this at present, and must rely on the empirical/observational results.

Finally, we comment on two other aspects of the PCA method for which a better understanding is now
available in light of the above analysis. First, Brunt et al (2003) found that PCA appears to operate
at first order -- i.e. in the case of an intermittent field when $\gamma_{1} \neq \gamma_{2}$, the 
PCA exponent $\alpha$ is better-correlated with the first-order index $\gamma_{1}$. This can be now understood
to be related to the ``phase-rolling'' effect (i.e. the $m$-dependent amplification of velocity fluctuations 
to roll the (co)sinusoid phase of the eigenimage structure) discussed above, which is a first order effect rather than
a root-mean-square effect. Second, it has been shown empirically that the recovered PCA exponent $\alpha$ is not strongly
affected by (column) density fluctuations (BH02, Roman-Duval et al 2011). While a full analysis of this effect is beyond
the scope of the current paper, a preliminary understanding of why this is
can be arrived at by considering the eigenimage structure given by equation~(\ref{i2approx}). An eigenimage of order $m$ is
the product of the column density ($m$-independent) and the (co)sinusoid ($m$-dependent). The ACF of such an eigenimage
is the Fourier transform of its power spectrum, which in turn is the square of its Fourier transform. A product in direct-space
transforms to a convolution in Fourier space, so the quantity of interest (the Fourier transform of the (co)sinusoid) is
convolved with the Fourier transform of the column density -- a function that is independent of order $m$. In the case of
uniform column density, this function is a delta function and the transform of the (co)sinusoid is unchanged. As column density
fluctuations become more important, a broadening of the column density transform is induced, but as long as this remains
narrow (in Fourier space) relative to the (co)sinusoid transform's Fourier-space width, no significant effect on the combined
power spectrum (and therefore ACF) will be induced. However, for an extremely variable column density field with a broad Fourier space
extent (as examined by Roman-Duval et al 2011) this must eventually break down. Roman-Duval et al 2010 determine that a density 
field with a lognormal PDF with $\sigma_{\ln(\rho/\rho_{0})} > 2$ is required for this to occur (see their Figure~5).

\section{Summary}

In this paper, we have derived and discussed analytic expressions for covariance matrices, eigenvectors, eigenvalues and
eigenimages expected from principal component analysis of molecular cloud emission lines, in the limit where these can
be represented by a collection of Gaussian line profiles with turbulent spatial correlations. Previous to this study,
the PCA method was based almost entirely on empirical analysis and lacked a firm theoretical basis.

We have derived an analytic calibration of the PCA method for measuring the spectrum of turbulent velocity fluctuations,
which agrees reasonably well with previous empirical calibrations. However, given the level of approximation in the analysis,
we see the analytic calibration more as a validation of the empirical calibration, rather than a replacement.  
We have also gained significant insight into the mechanisms by which PCA operates, allowing us to explain more esoteric 
aspects of the method, such as its preferential operation at first order and its general robustness against (column) density fluctuations.

\section*{Acknowledgements}
C.~B. is funded in part by the UK Science and Technology Facilities Council grant ST/J001627/1 
(``From Molecular Clouds to Exoplanets'') and the ERC grant ERC-2011-StG\_20101014 (``LOCALSTAR''),
both held at the University of Exeter. We thank the referee, Erik Rosolowsky, for a very perceptive
and thorough review and sensible suggestions that improved the clarity of the text.


\label{lastpage}


\begin{thebibliography}{99}
\bibitem[\protect\citeauthoryear{Brunt}{2003}]{b03b} Brunt, C. M., 2003(a), ApJ, 583, 280
\bibitem[\protect\citeauthoryear{Brunt}{2003}]{b03a} Brunt, C. M., 2003(b), ApJ, 584, 293
\bibitem[\protect\citeauthoryear{Brunt \& Heyer}{2002}]{bh02a} Brunt, C.~M., \& Heyer, M.~H., 2002(a), ApJ, 566, 276 (BH02)
\bibitem[\protect\citeauthoryear{Brunt \& Heyer}{2002}]{bh02b} Brunt, C.~M., \& Heyer, M.~H., 2002(b), ApJ, 566, 289
\bibitem[\protect\citeauthoryear{Brunt et al}{2003}]{bhvsb03} Brunt, C.~M., Heyer, M.~H., V\'{a}zquez-Semadeni, E., \& Pichardo, B., 2003, ApJ, 595, 824
\bibitem[\protect\citeauthoryear{Brunt, Heyer, \& Mac Low}{2009}]{bhml09} Brunt, C. M., Heyer, M. H., \& Mac Low, 2009, A{\&}A, 504, 883
\bibitem[\protect\citeauthoryear{Brunt \& Mac Low}{2004}]{bml04} Brunt, C. M., \& Mac Low, 2004, ApJ, 604, 196 
\bibitem[\protect\citeauthoryear{Heyer \& Brunt}{2004}]{hb04} Heyer, M.~H., \& Brunt, C.~M., 2004, ApJL, 615, 45 
\bibitem[\protect\citeauthoryear{Heyer et al}{2008}]{heyer2008} Heyer, M.~H., Gong, H., Ostriker, E.  \& Brunt, C.~M., 2008,  ApJ, 680, 420
\bibitem[\protect\citeauthoryear{Heyer \& Brunt}{2012}]{hb12} Heyer, M.~H., \& Brunt, C.~M., 2012, MNRAS, 420, 1562 
\bibitem[\protect\citeauthoryear{Heyer \& Schloerb}{1997}]{hs97} Heyer, M.~H., \& Schloerb, F.~P., 1997, ApJ, 475, 173 (HS97)
\bibitem[\protect\citeauthoryear{Heyer et al}{1998}]{ogs98} Heyer, M.~H., Brunt, C.~M., Snell, R.~L, Howe, J.~E., Schloerb, F.~P., \& Carpenter, J.~M., 1998, ApJS, 115, 241
\bibitem[\protect\citeauthoryear{Kleiner \& Dickman}{1985}]{kd85} Kleiner, S.~C., \& Dickman, R.~L., 1985, 295, 466
\bibitem[\protect\citeauthoryear{Lazarian \& Pogosyan}{2000}]{lp00} Lazarian, A., \& Pogosyan, D., 2000, ApJ, 537, 720
\bibitem[\protect\citeauthoryear{Miesch \& Bally}{1994}]{mb94} Miesch, M.~S., \& Bally, J., ApJ, 1994, 429, 645
\bibitem[\protect\citeauthoryear{Roman-Duvval et al}{2011}]{rd11} Roman-Duval, J., Federrath, C., Brunt, C.~M., Heyer, M.~H., Jackson, J.~M., \& Klessen, R.~S., 2011, ApJ, 740, 120
\bibitem[\protect\citeauthoryear{Scalo}{1984}]{s84} Scalo, J.~M., 1984, ApJ, 277, 556 
\bibitem[\protect\citeauthoryear{Stutzki et al}{1998}]{s98} Stutzki, J., Bensch, F., Heithausen, A., Ossenkopf, V., \& Zeilinsky, M., 1998, A{\&}A, 336, 697
\end{thebibliography}
\end{document}